\documentclass[10pt,twocolumn]{article}

\usepackage[a4paper,top=1.65cm,bottom=1.65cm,left=1.55cm,right=1.55cm,columnsep=0.65cm]{geometry}
\usepackage{newtxtext,newtxmath}
\usepackage{microtype}
\usepackage{graphicx}
\usepackage{booktabs}
\usepackage{siunitx}
\usepackage{cite}
\usepackage{url}
\usepackage[hidelinks]{hyperref}
\usepackage{caption}
\usepackage{enumitem}
\usepackage{balance}
\usepackage{fancyhdr}

\newcommand{\acceptednote}{%
    \scriptsize
    The present article has been accepted by The Physics Teacher -- AIP.\\
    After it is published, it will be found at
    \url{https://doi.org/10.1119/5.0331945}%
}

\setlist{nosep}

\fancypagestyle{plain}{%
    \fancyhf{}
    \fancyfoot[C]{\acceptednote}
}

\title{\vspace{-1.2em}\bfseries\Large Measuring the Speed of Light with a Time-of-Flight Setup:\\[2pt] An Authentic Experiment for High School}
\author{\textbf{Alberto Mazzi}\\
Istituto ``A. Degasperi'', Provincia Autonoma di Trento,
Borgo Valsugana, TN, Italy}
\date{}

\begin{document}
\twocolumn[\maketitle\vspace{-1.4em}]

Students often perceive scientific experiments as clean, straightforward procedures with predictable outcomes. Students' beliefs about the nature of experimental physics typically remain inconsistent through the initial stages of undergraduate education.~\cite{wilcox2017,jacobs2017} In reality, experimental work in physics involves hands-on activities such as soldering circuits, assembling provisional setups, and building functional mock-ups from simple materials before reliable measurements can be achieved. To address this gap, we designed a laboratory activity in which students were completely involved in the experiment design, construction and operation with the final goal of measuring the speed of light with a low-cost and compact setup. The educational objective of this activity was double: firstly, to make students aware of the importance of experimental choices and constraints in the setup design and secondly, to discover the potentiality of semiconductor technologies. As far back as the 1990s, with the spread of compact disk players,~\cite{yoshikawa1995,armstrong1992} laser diodes and photodiodes capable of fast timing performance in the order of nanoseconds became largely available. In the last decades, consumer electronic devices made high performance technologies ubiquitous. Nevertheless, the idea of measuring ultrashort time intervals and the relevance of such measurements in science and technology remain unknown to most students. Nowadays, digital instruments with significant performance such as waveform generators and oscilloscopes are affordable for high school physics labs, paving the way to a closer contact with the world of experimental physics.~\cite{bochkov2016}

\section{Design and construction}
The laboratory experiment presented in this article was realized in a physics course in an Italian secondary school. Some preliminary notions needed for the understanding of the phenomenology involved in the experiment were discussed in class. In particular, in addition to the classic introduction to linear electric circuits, the behavior of diodes was presented. Particular emphasis was given to junction formation and bias operating conditions, though at an introductory level.

The experiment was designed in a collaborative way, with the active participation of students. The baseline idea was a compact and low-cost setup for measuring the speed of light. Considering the instrumentation available in our physics laboratory, we opted for a time-of-flight method, where a light source is operated intermittently and the light output is detected at a variable distance. For the sake of simplicity, the light path was assumed to be in air. The need for compromises in the experiment design arose naturally. On one hand, in the case of a short optical path the time resolution of electronic instrumentation is critical but, on the other hand, for longer paths signal integrity may become a problem.

\begin{figure}[t]
\centering
\includegraphics[width=\columnwidth]{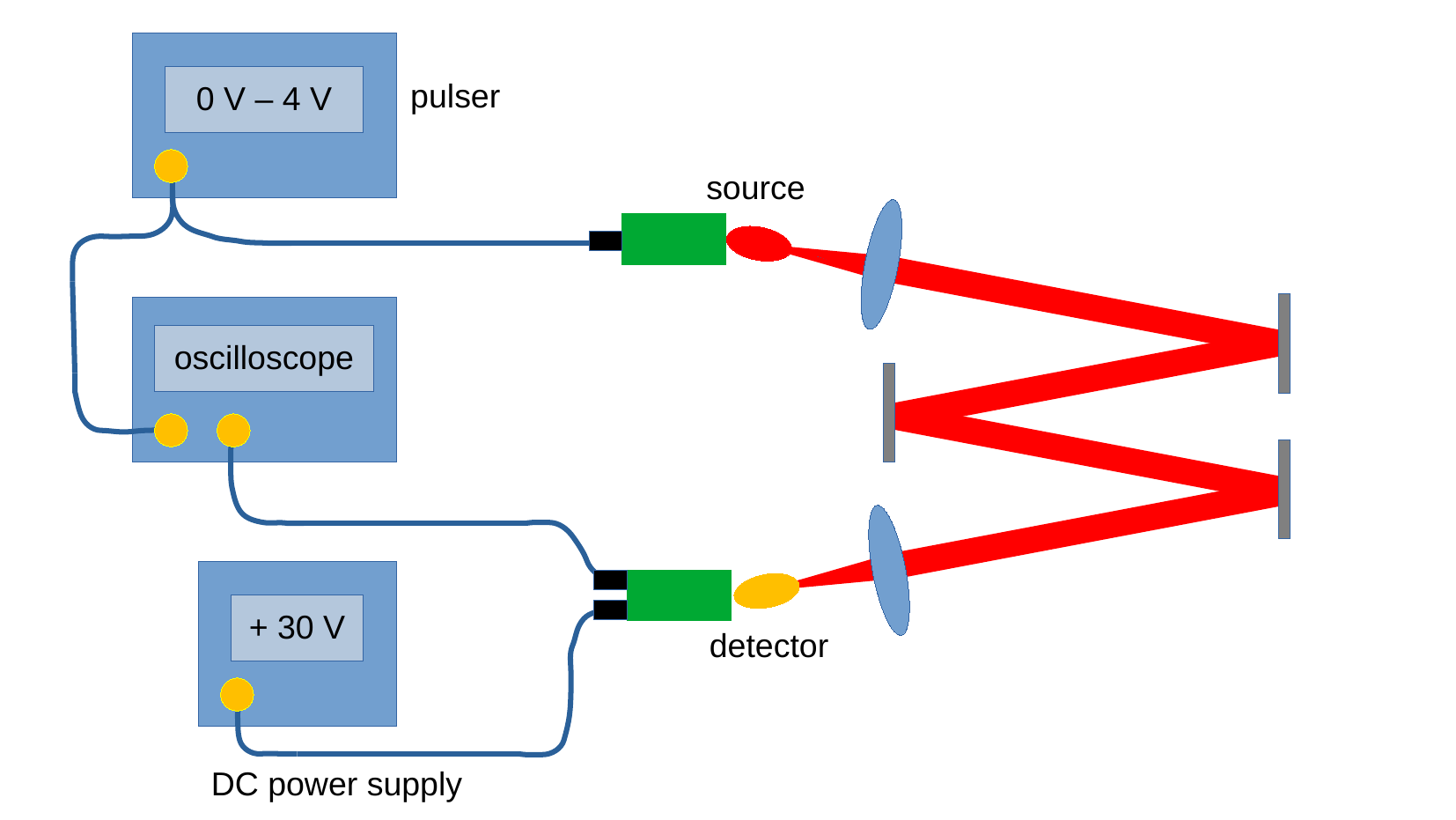}
\caption{Apparatus schematics. All wirings represented in the scheme are 50 cm long coaxial cables.}
\label{fig:apparatus}
\end{figure}

The electronic setup was based on two instruments already available in our laboratory: a Siglent SDG 1032X digital waveform generator with a maximum output frequency of 30 MHz and a Siglent SDS 1202X-E 2-channel digital oscilloscope with 200 MHz bandwidth. These instruments set a lower limit for the final experiment time resolution: in particular the most limiting factor was the minimum rise time of the signal generator, of about 4 ns in the square waveform mode. As a light source we initially considered some common visible light emitting diodes (LEDs), but from preliminary measurements we realized that they featured a slow time response, typically in the order of 100 ns, thus requiring an optical path of more than 30 m in order to obtain a detectable time delay. To overcome this limit, an infrared laser diode with fast response was utilized (TT Electronics OPV332 with peak emission at 850 nm, rise time 0.1 ns). As a photodetector we opted for a broad spectral sensitivity PIN silicon photodiode (TT Electronics OP 906, with typical rise time of 5 ns at 20 V reverse bias). Based on these instruments, in the initial stages, we approximated the time resolution of the apparatus to be below 10 ns. As a consequence, the minimum needed optical path was about 3 m.

The optical system used in our experiment is aimed at minimizing the footprint in order to arrange all the instruments on a common lab bench. A first biconvex lens is placed at focal distance from the light source to produce a parallel light output. The light beam is then reflected on three flat mirrors before reaching the receiver. A second biconvex lens is used to focus the light beam on the photodiode. The choice of lens dimensions and focal length required some trade-off: with a short focal length high mechanical precision is needed in the placement of emitter or receiver at the focal point, while a large focal length potentially resulted in light intensity loss depending on the lens diameter. The students performed some preliminary tests with different lenses available in our lab by using a red LED as a light source. They demonstrated that, by using a biconvex lens with 15 cm focal length and 5 cm diameter it was possible to obtain a reasonably good signal intensity at a distance of at least 10 m. The electrical connections and the conceptual optical schemes are reported in Fig.~\ref{fig:apparatus}.

\begin{figure*}[!t]
\centering
\includegraphics[width=0.92\textwidth]{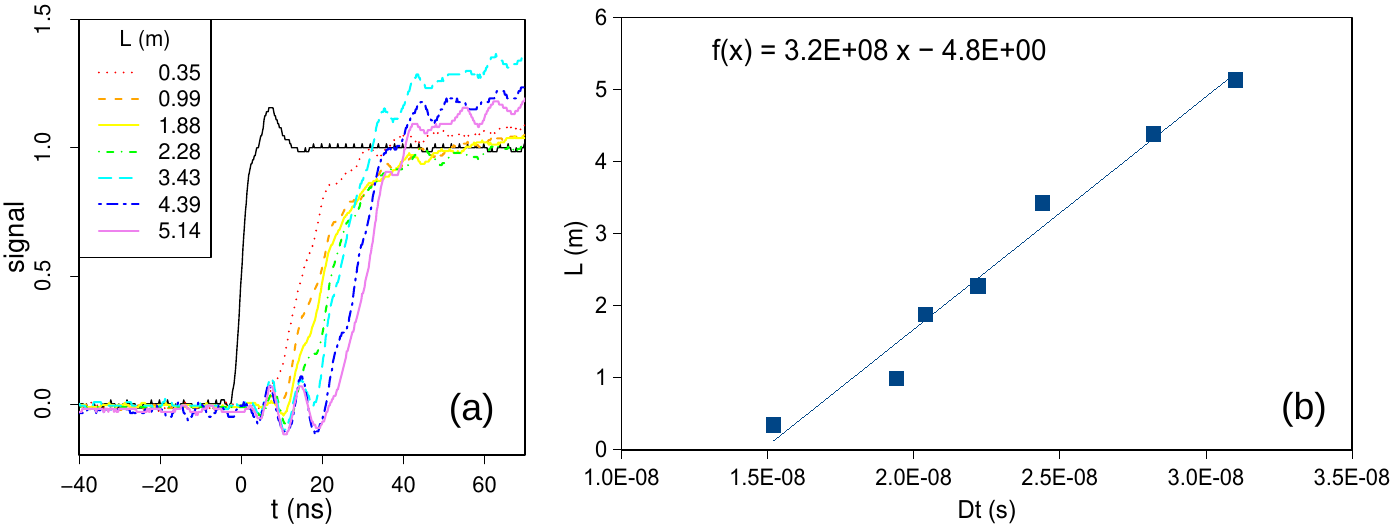}
\caption{(a) Oscilloscope traces: laser diode signal (black) and photodiode response (colored traces) for different optical path lengths. (b) Optical path length as a function of time delay.}
\label{fig:traces}
\end{figure*}

The electronic boards for laser diode and photodiode operation were assembled by the students. The circuits were designed as simple as possible, thus allowing students to understand the physics of device operation. The laser diode was mounted on a stripboard through a specific socket and it received the output signal of waveform generator through a coaxial cable. The power signal was a square waveform with a high level of 4.0 V in forward bias, low level of 0 V, frequency of 1 kHz and duty cycle of 10\%. The photodiode was mounted on a similar stripboard, in series with a 68~$\Omega$ resistance that was used for photocurrent readout. The photodiode was biased in reverse mode at 30.0 V DC. The high bias voltage allows increasing junction depletion thus reducing diode capacitance.

The experimental setup was operated in two steps: once the source-detector distance was set, the system was aligned by using a visible LED, then the LED was replaced with the IR laser diode. With the laser diode mounted, the photocurrent signal measured through the oscilloscope was maximized with a fine optical adjustment. The experiment was performed in ambient light, but paying attention to avoid eye exposure to collimated light.

The overall time commitment for the design, construction and operation of the experiment was about ten hours, spread over two months, as needed for purchasing the required components and arranging all the materials.

\section{Results and discussion}
In the final configuration of the apparatus, the time delay between emission and detection of light pulses was measured for seven different lengths of the optical path. The waveforms received by the oscilloscope on the two channels were recorded to be analyzed offline. A simple code written for RStudio~\cite{simoneau2015} environment allowed to precisely discriminate the time delay. In order to obtain an optimal time resolution, a voltage threshold was set at half height between the low level and high level, both for the bias signal of the laser diode and for the photodiode signal. The waveforms were collected in average mode, so that the acquired signals are the mean behavior of 4 subsequent light pulses. This reduces the random noise at the high level and low level, while reduction of time jitter at the threshold level was not appreciable, since the signal edge was typically stable and sharp.

The oscilloscope traces shown in Fig.~\ref{fig:traces} (a) illustrate the typical temporal evolution of the signals, where the black line represents the pulser signal, while the colored lines correspond to the photodiode responses at different optical path lengths.

\begin{figure}[!t]
\centering
\includegraphics[angle=90,width=0.95\columnwidth]{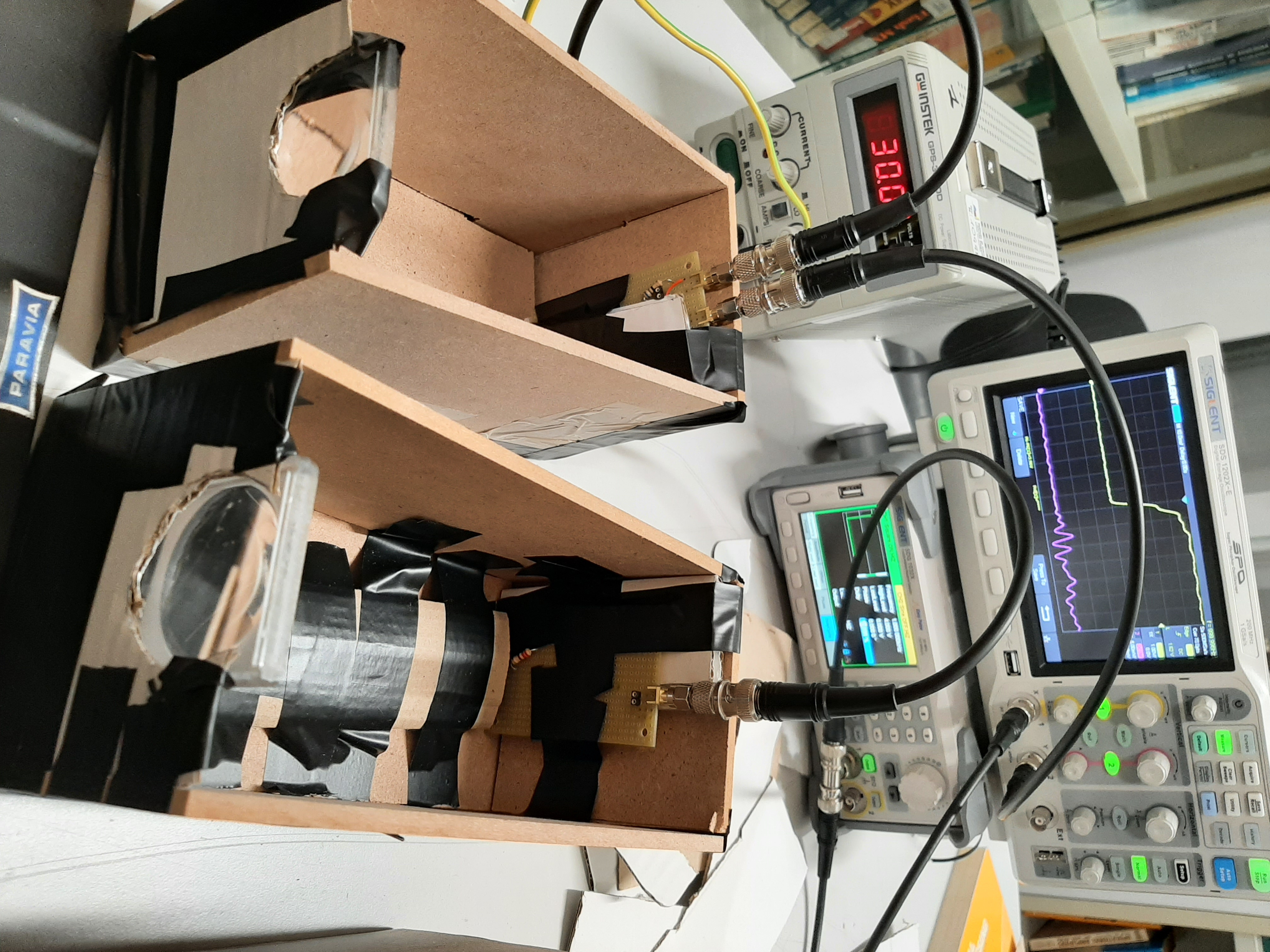}
\caption{A detail of the experimental setup. In the foreground, the two cases for the laser diode and photodiode. In the background, the DC power supply, the oscilloscope and the waveform generator.}
\label{fig:setup}
\end{figure}

The time origin is set at the laser diode threshold, while the vertical axis is normalized to a unitary steady-state amplitude reached after some transient oscillations. The measurements presented in Fig.~\ref{fig:traces} (b) show a linear dependence of time delay and source-detector distance. The slope of the interpolating line was estimated as $(3.2\pm0.2)\times10^{8}$~m/s, compatible with the expected value of the speed of light in air. The experimental uncertainty was evaluated as the standard error of the interpolation, performed through LibreOffice Calc. The non-zero intercept is due to a systematic time delay of $(15\pm2)$~ns. This aspect depends on the combination of several contributions: the rise time of the laser diode and photodiode and the signal propagation from the photodiode to the oscilloscope. A thorough evaluation of experimental uncertainty would be overly complex for the time delay values. The uncertainty in the length measurements ranges from approximately 1 to 4 cm, increasing with the number of reflections, and can be considered negligible compared with timing uncertainty. On the other hand, the use of spreadsheet functions for uncertainty estimation could provide a valuable opportunity to discuss with students the basic principles of least-squares fitting.

A similar approach for the measurement of the speed of light at a high school level was used more than twenty years ago~\cite{mak2000}, but with significant differences. In that case, the oscilloscope was a cathode ray device with a sweeping frequency of 10--20 MHz and a sinusoidal waveform was used to modulate a laser pointer. The speed of light was measured with a phase shift method by using an optical path of about 12 m. As we demonstrated, with technology that is nowadays affordable for a high school physics laboratory, the equipment and the measurement method can be simplified and made more compact. By using simple components such as laser diodes and photodiodes in T-1 package that can be conveniently mounted on stripboards, the students can be involved in the design and fabrication of the experiment. A partial view of the hand-made equipment is shown in Fig.~\ref{fig:setup}.

In modern metrology, the speed of light in vacuum is defined as an exact conventional value ($299\,792\,458$~m/s). From a rigorous standpoint, the activity described here can therefore be viewed as a measurement of the refractive index of air or as a distance measurement based on the time of flight of a light pulse. Such techniques are commonly employed in light detection and ranging (LIDAR) systems used in industrial, automotive, and satellite contexts. However, from an educational perspective, students found the idea of measuring the speed of light to be significantly more motivating than interpreting the experiment as a distance-measurement method.

\section*{Acknowledgments}
The activity presented in this article was developed in the framework of a competition for high school students organized by the Cherenkov Telescope Array Plus program and cosponsored by the Italian National Institute of Astrophysics (INAF) and the Italian National Institute for Nuclear Physics (INFN). The competition was financed by the Italy's Recovery and Resilience Plan, coordinated by the Italian Ministry of Education and Merit (MIM) and European Union -- NextGenerationEU. The author is grateful to the students who participated in the project. Their enthusiasm was a strong motivation for this work.

\balance

\end{document}